\documentclass{article}
\usepackage[margin=1in]{geometry}
\usepackage{amsmath, amsfonts, graphicx, authblk}

\newcommand{\C}{\mathbb{C}} 
\newcommand{\R}{\mathbb{R}} 
\newcommand{\Z}{\mathbb{Z}} 
\newcommand{\im}{i} 
\newcommand{\graph}{\mathcal{G}} 
\newcommand{\V}{\mathcal{V}} 
\newcommand{\edges}{\mathcal{E}} 
\newcommand{\T}{[0,1]} 
\newcommand{\ZV}{\Z\times\V} 
\newcommand{\TV}{\T\times\V} 

\newcommand{\defined}{:=} 
\newcommand{\conv}{\ast} 
\newcommand{\intT}{\int_0^1} 

\newcommand{\st}{~|~} 
\newcommand{\ie}{\textit{i.e.} } 
\newcommand{\eg}{\textit{e.g.} } 
\newcommand{\etal}{\textit{et al.} } 

\newcommand{\abs}[1]{\left| #1 \right|} 
\newcommand{\set}[1]{\left\{ #1 \right\}} 
\newcommand{\tens}[1]{\mathbf{#1}} 
\newcommand{\vect}[1]{\mathbf{#1}} 
\newcommand{\Oh}[1]{\mathcal{O}\left(#1\right)} 
\newcommand{\exparg}[1]{\exp\left\{#1\right\}} 
\newcommand{\earg}[1]{e^{#1}} 
\newcommand{\lp}[2]{\ell^{#2}\left(#1\right)} 
\newcommand{\Lp}[2]{L^{#2}\left(#1\right)} 
\newcommand{\norm}[1]{\left\| #1 \right\|} 
\newcommand{\lpnorm}[2]{\norm{#1}_{\lp{#2}{2}}}
\newcommand{\Lpnorm}[2]{\norm{#1}_{\Lp{#2}{2}}} 

\newcommand{\hilbert}{\lp{\ZV}{2}} 
\newcommand{\dual}{\Lp{\TV}{2}} 

\newcommand{\identity}{\tens{I}} 
\newcommand{\kernel}{\tens{K}} 
\newcommand{\shift}{\tens{S}} 
\newcommand{\proj}{\tens{P}} 
\newcommand{\nil}{\tens{N}} 
\newcommand{\F}{\mathcal{F}} 
\newcommand{\iF}{\mathcal{F}^*} 

\newcommand{\bnorm}[2]{\norm{#1}_{\mathcal{B}\left(#2\right)}} 
\newcommand{\LB}[1]{\mathcal{B}\left( #1 \right)} 
\newcommand{\spectrum}[1]{\Lambda\left(#1\right)} 

\title{Learning flexible representations of stochastic processes on graphs}

\author[1,3]{Addison W. Bohannon}
\author[1]{Brian M. Sadler}
\author[2,3]{Radu V. Balan}
\affil[1]{US Army Research Laboratory, Adelphi, MD}
\affil[2]{Department of Mathematics, University of Maryland, College Park, MD}
\affil[3]{Center for Sci. Comp. and Mathematical Modeling, University of Maryland, College Park, MD}

\date{}

\begin{document}
\maketitle
\begin{abstract}
Graph convolutional networks adapt the architecture of convolutional neural networks to learn rich representations of data supported on arbitrary graphs by replacing the convolution operations of convolutional neural networks with graph-dependent linear operations. However, these graph-dependent linear operations are developed for scalar functions supported on undirected graphs. We propose a class of linear operations for stochastic (time-varying) processes on directed (or undirected) graphs to be used in graph convolutional networks. We propose a parameterization of such linear operations using functional calculus to achieve arbitrarily low learning complexity. The proposed approach is shown to model richer behaviors and display greater flexibility in learning representations than product graph methods.
\end{abstract}

\section{Introduction}\label{sec:introduction}

The large amounts of data and rich interactions characteristic of complex networks such as those observed in brain imaging and social networks motivate the need for rich representations for use in learning. Convolutional neural networks offer a means of learning rich representations of data by composing convolutions, pooling, and nonlinear activation functions \cite{lecun2015deep}. However, in Bruna \etal \cite{bruna_spectral_2014}, it is argued that the success of convolutional neural networks for images, video, and speech can be attributed to the special statistical properties of these domains (\ie sampled on a regular grid; local similarity, translation invariance, and multi-scale structure) and does not generalize to data with arbitrary graph structure. 

This motivates a generalization of the convolution to data supported on an arbitrary graph, and in Bruna \etal \cite{bruna_spectral_2014}, a graph-specific convolution is defined consistent with the graph Fourier transform proposed in Shuman \etal \cite{shuman_emerging_2013}. In this formulation, the eigenvectors of the graph Laplacian form the basis for linear operations as well as imparting the topology of the graph. A learning algorithm then optimizes the eigenvalues of the linear operator. Application of this learning representation in so-called graph convolutional networks has yielded state-of-the-art results in applications such as network analysis, computer graphics, and medical imaging \cite{bronstein_geometric_2017}. 

However, as identified in Bronstein \etal \cite{bronstein_geometric_2017}, this procedure has limitations, and this chapter aims to address two of them. First, it is not clear how to apply these techniques to stochastic (time-varying) processes on graphs, and second, the graph signal processing approach advocated by Shuman \etal \cite{shuman_emerging_2013} works only for undirected graphs. These limitations severely reduce the domains in which such learning representations can apply. This chapter aims to address these two gaps by proposing a theoretical framework for designing and learning graph-specific linear operators that act on stochastic processes on directed (or undirected) graphs. Together with pooling operations and nonlinear activation functions, it is hypothesized that such linear operators would lead to learning rich representations of stochastic processes on graphs. Throughout, consideration is given to learning complexity.

The paper proceeds as follows. Section \ref{sec:related_work} discusses work related to filtering and linear modeling of stochastic processes supported on graphs. Section \ref{sec:preliminaries} establishes some preliminary notation and theory from harmonic analysis. Sec. \ref{sec:covariant} motivates the learning of covariant linear operations and proposes the theoretical framework for designing them. Section \ref{sec:covariant_func_calc} proposes the use of functional calculus to design and learn covariant representations with arbitrarily low complexity. Section \ref{sec:example} then compares the proposed approach to an alternative approach (Sandryhaila and Moura \cite{sandryhaila_big_2014}) for an example problem.

\section{Related Work}\label{sec:related_work}

Learning representations for graph structured data has featured in two recent review articles, that of Bronstein \etal \cite{bronstein_geometric_2017} and Hamilton \etal \cite{hamilton_representation_2017}. These reviews discuss defining convolutional neural networks with graph-specific linear operators as in Bruna \etal \cite{bruna_spectral_2014}. In Bruna \etal \cite{bruna_spectral_2014}, the learnable parameters comprise the eigenvalues of a linear operator with eigenvectors fixed by the graph Laplacian. In Defferrard \etal \cite{defferrard_convolutional_2016} and Kipf and Welling \cite{kipf_semi-supervised_2017}, the learnable parameters are instead the coefficients of a polynomial on the graph Laplacian, reducing the learning complexity and leading to superior results in application.

Graph signal processing as proposed in Sandryhaila and Moura \cite{sandryhaila_discrete_2013, sandryhaila_discrete_2014} and Shuman \etal \cite{shuman_emerging_2013} extends the traditional tools of time-series signal processing to scalar functions supported on the nodes of a graph. Theoretical extensions for transforms, sampling, and filtering have been established and applied in various domains. For a recent review, see Ortega \etal \cite{ortega_graph_2017}. Analysis of stochastic processes supported on a graph is a special case of graph signal processing first addressed in Sandryhaila and Moura \cite{sandryhaila_big_2014}. In this work, the authors propose a generalization of their graph signal processing approach to multi-variate observations which can be modeled in a factor graph. This same idea underlies the work of Loukas and Foucard \cite{loukas_frequency_2016} and Grassi \etal \cite{grassi_time-vertex_2018} which specifically address time-varying graph signals.

\section{Preliminaries}\label{sec:preliminaries}

Let $\graph=\left(\V,\edges\right)$ be a graph with nodes $\V=\set{0,\ldots,n-1}$ and edges $\edges\subseteq\V\times\V$. We consider stochastic processes which take values on $\V$, indexed by time in $\Z$, \ie a sequence of vector-valued functions of $\V$, $\vect{x}=\set{\vect{x}[t]:\V\to\C^n}_{t\in\Z}$. The image of this function should be thought of as representing an attribute of the vertices of the graph at the indexed time. That attribute could be the action of posting or liking a message by an individual in a social network or the recorded activity at an electrode or brain region. 

We consider a particular subset of stochastic processes on this graph, those which are square summable,
\begin{equation}\label{eq:hilbert}
\hilbert = \set{ \set{\vect{x}[t]\in\C^n}_{t\in\Z} \st \lpnorm{\vect{x}}{\ZV}<\infty }.
\end{equation}
Here, $\lpnorm{\vect{x}}{\ZV}=\sum_{t\in\Z} \norm{\vect{x}[t]}_2^2$.
We want to find generalizations of the convolution on these functions, that is to say bounded linear transformations, $\tens{A}:\hilbert\to\hilbert$. A linear transformation is called bounded if
\begin{equation}\label{eq:bnorm}
\bnorm{\tens{A}}{\hilbert} = \sup_{ \lpnorm{\vect{x}}{\ZV}=1 } \sum_{t\in\Z} \norm{\left(\tens{A}\vect{x}\right)[t]}_2^2 < \infty.
\end{equation}
Bounded linear transformations are denoted $\LB{\hilbert}$. Although $\tens{A}\in\LB{\hilbert}$ maps infinite vector-valued sequences into infinite vector-valued sequences, its action on $\hilbert$ is similar to a matrix-vector product. For any $\tens{A}\in\LB{\hilbert}$, there exists a unique kernel function $\kernel:\Z\times\Z\to\C^{n\times n}$ such that 
\begin{equation}\label{eq:action_of_A_kernel}
\left(\tens{A}\vect{x}\right)[t] = \lim_{N\to\infty} \left(\sum_{s=-N}^N \kernel(t,s) \vect{x}[s]\right).
\end{equation}

More specifically, we consider bounded linear transformations which have a Laurent structure, those for which $\kernel(t,s)=\kernel(t+d,s+d)$ for all $d\in\Z$. This gives $\tens{A}$ a bi-infinite block Toeplitz structure,
\begin{equation}\label{eq:matrix_vector_representation}
\tens{A}\vect{x} = \begin{bmatrix}
\ddots & \ddots & \ddots \\
\ddots & \kernel_0 & \kernel_{-1} & \kernel_{-2} \\
\ddots & \kernel_1 & \kernel_ 0 & \kernel_{-1} & \ddots \\
& \kernel_2 & \kernel_1 & \kernel_0 & \ddots \\
& & \ddots & \ddots & \ddots
\end{bmatrix} \begin{bmatrix}
\vdots \\ \vect{x}[-1] \\ \vect{x}[0] \\ \vect{x}[1] \\ \vdots
\end{bmatrix}
\end{equation}
where $\kernel_t=\kernel(t,0)$. A Laurent operator is seen to be a generalization of the convolution on $\lp{\Z}{2}$, where for $x,y\in\lp{\Z}{2}$, $(x\conv y)[t] = \sum_{s\in\Z} x[t-s] y[s]$. Much like convolution is diagonalized by the Fourier transform, Laurent operators act multiplicatively after a Fourier transform. For $\vect{x}\in\hilbert$, we define the Fourier transform as
\begin{equation}\label{eq:fourier_transform}
\left(\F \vect{x}\right)(\omega) \defined \sum_{t\in\Z} \earg{2\pi\im\omega t} \vect{x}[t].
\end{equation}
Then, for a $\tens{A}\in\LB{\hilbert}$ Laurent and $\vect{x}\in\hilbert$,
\begin{equation}\label{eq:A_diagonalized_by_F}
\left( \F \tens{A} \vect{x} \right)(\omega) = \hat{\tens{A}}(\omega) \cdot \hat{\vect{x}}(\omega)
\end{equation}
where
\begin{equation}\label{eq:A_spectral}
\hat{\tens{A}}(\omega) \defined \sum_{t\in\Z} \earg{2\pi\im\omega t} \kernel_t .
\end{equation}

We make our analysis complete by defining
\begin{equation}\label{eq:dual}
\dual = \set{ \hat{\vect{x}}: \T\to\C^n \st \Lpnorm{\hat{\vect{x}}}{\TV}<\infty },
\end{equation}
where $\Lpnorm{\hat{\vect{x}}}{\TV}=\intT \norm{\hat{\vect{x}}(\omega)}_2^2 d\omega$. The associated Fourier transform for $\hat{\vect{x}}\in\dual$ is
\begin{equation}\label{eq:inverse_fourier_transform}
\left(\iF \hat{\vect{x}}\right)[t] \defined \intT \earg{-2\pi\im\omega t} \hat{\vect{x}}(\omega) d\omega.
\end{equation}
Now, we can say that $\F:\hilbert\to\dual$ is bijective and unitary, and the same for $\iF:\dual\to\hilbert$. Moreover, $\F\iF=\iF\F=\identity$. 

\section{Learning robust representations of $\hilbert$}\label{sec:covariant}

The goal of this paper is to present a framework for learning robust representations of stochastic processes on graphs. Important to learning and generalization is invariance or covariance to particular group actions. For example, convolutional neural networks depend on the covariance of convolution to translation and invariance of pooling to small deformations \cite{mallat_group_2012, mallat_understanding_2016}. This motivates defining linear operators on $\hilbert$ with such symmetries. 

Here, we consider covariance to an arbitrary group generator, $\shift\in\LB{\hilbert}$ Laurent,
\begin{equation}\label{eq:covariance}
\left( \tens{A}\shift\vect{x} \right)[t] = \left( \shift\tens{A}\vect{x} \right)[t]
\end{equation}
where $\vect{x}\in\hilbert$. $\shift$ could be thought of as the weighted adjacency matrix (as in Sandryhaila and Moura \cite{sandryhaila_discrete_2013}) or the Laplacian of $\graph$ (as in Shuman \etal \cite{shuman_emerging_2013}) with an additional dimension of time. Thus, $[\kernel_t]_{j,k}$ for $j,k\in\V$ where $\kernel$ is the kernel function of $\shift$ can be understood as the weighted edge between nodes $j$ and $k$ at a temporal distance of $t$. Moreover, it defines the mechanism by which nodes interact in time and space. 

For a given $\shift$, we want to find a parameterization of $\tens{A}\in\LB{\hilbert}$ to satisfy Eq. \eqref{eq:covariance}. Given a parameterization of $\tens{A}$ and a model which depends on $\tens{A}$, the learning problem is to estimate the parameters of $\tens{A}$ conditioned on observed data. The learning complexity then depends on the parameterization of $\tens{A}$.

%
%

Let $\shift$ have kernel function $\kernel$ as in Eq. \eqref{eq:action_of_A_kernel}. Since $\shift$ is Laurent, it admits a frequency representation $\hat{\shift}:\T\to\C^{n\times n}$ as in Eq. \eqref{eq:A_spectral}. Further, pointwise for $\omega\in\T$, we have
\begin{equation}\label{eq:jordan}
\hat{\shift}(\omega) = \sum_{k=0}^{m(\omega)} \lambda_k(\omega) \proj_k(\omega) + \nil_k(\omega)
\end{equation}
where $0<m(\omega)\le n$ and the following conditions hold for all $j,k=\set{0,\ldots,m(\omega)}$ and $\omega\in\T$.
\begin{enumerate}
\item $\sum_{k=0}^{m(\omega)} \proj_k(\omega)=\identity$
\item $\proj_k(\omega)\proj_j(\omega)=\proj_j(\omega)\proj_k(\omega)=\delta_{jk}\proj_k(\omega)$
\item $\nil_k(\omega)=\proj_k(\omega)\nil_k(\omega)\proj_k(\omega)$
\item $\left(\nil_k(\omega)\right)^n=\tens{0}$
\end{enumerate}
Eq. \eqref{eq:jordan} is known as the Jordan spectral representation, and it is unique \cite{simon2015operator}. We can use this result to parameterize $\tens{A}$ in the frequency domain because Eq. \eqref{eq:covariance} is equivalent to $\hat{\tens{A}}(\omega)\cdot\hat{\shift}(\omega) = \hat{\shift}(\omega)\cdot\hat{\tens{A}}(\omega)$ almost everywhere on $\omega\in\T$. Therefore, if 
\begin{equation}\label{eq:A_covariant_to_S}
\hat{\tens{A}}(\omega) = \sum_{k=0}^{m(\omega)} \hat{a}_k(\omega) \proj_k(\omega) + \nil_k(\omega)
\end{equation}
for $\hat{a}_k\in\Lp{\T}{\infty}$ for $k\in\V$, then $\tens{A}\in\LB{\hilbert}$ satisfies Eq. \eqref{eq:covariance}.

As a function of $\omega$, there is very little that can be said about the multiplicity of the eigenvalues (\ie $m(\omega)$) and the associated invariant subspaces (\ie $\proj_k(\omega)$ and $\nil_k(\omega)$) for an arbitrary $\shift\in\LB{\hilbert}$. Much stronger results exist for $\hat{\shift}$ holomorphic on an annulus, $\set{z\in\C\st 1-\epsilon<\abs{z}<1+\epsilon}$ for $\epsilon>0$ which is satisfied for $\norm{\kernel_t}_2<C_1(1+\epsilon)^{-t}$ for $t>0$ and $\norm{\kernel_t}_2<C_2(1-\epsilon)^t$ for $t<0$ and constants $C_1,C_2>0$. Such a restriction is satisfied by assuming that interaction between nodes beyond a sufficient temporal distance is negligible. We assume from now on that $\hat{\shift}:\T\to\C^{n\times n}$ is indeed a holomorphic matrix-valued function on an appropriate annulus. Then, $\lambda_k(\omega)$, $\proj_k(\omega)$, and $\nil_k(\omega)$ are holomorphic functions for all $k=\set{0,\ldots, m}$ and $\omega\in\T$. Moreover, $m(\omega)=m$ almost everywhere on $\omega\in\T$ (see \cite{kato1966perturbation} for a full discussion of analytic perturbation theory).

With these assumptions, a linear operation that is covariant to an arbitrary graph structure $\shift$ is defined by $\Oh{m}$ parameters (where $m$ scales with $n$), $\hat{a}_k\in\Lp{\T}{\infty}$ for $k\in\set{0,\ldots,m}$. 
\begin{equation}\label{eq:action_graph_covariant}
\left(\tens{A}\vect{x}\right)[t] = \left( \iF \left[ \sum_{k=0}^m \hat{a}_k(\omega) \proj_k(\omega) + \nil_k(\omega) \right] \cdot \hat{\vect{x}}(\omega) \right)[t].
\end{equation}
This result generalizes the spectral construction (Eq. 3.2) of Bruna \etal \cite{bruna_spectral_2014}, where now $\vect{x}$ is a function of time.

\section{Learning robust representations of $\hilbert$ with arbitrarily low complexity}\label{sec:covariant_func_calc}

Our learning framework entails defining parameterizations of $\tens{A}$ which satisfy Eq. \eqref{eq:covariance} for some $\shift\in\LB{\hilbert}$ and estimating the parameters of $\tens{A}$ conditioned on data observations. By Eq. \eqref{eq:action_graph_covariant}, the learning problem has complexity $\Oh{m}$, which is linear in the size of $\V$. Ideally, we want a parameterization of $\tens{A}$ that leads to sublinear learning complexity as is achieved for compactly supported convolutions. Defferrard \etal \cite{defferrard_convolutional_2016} propose for the scalar graph signal case to learn polynomials of the graph Laplacian instead of spectral multipliers as in Bruna \etal \cite{bruna_spectral_2014}. The corollary to our framework would be polynomials of $\shift$, \ie $\tens{A}=\sum_{k=0}^{p} a_k \shift^k$ for $0\le p<n$, which results in $p$ learnable parameters, $\set{a_k}_{k=0}^p$, and necessarily satisfies Eq. \eqref{eq:covariance}. That we can define linear transformations by polynomials is a special case of a more comprehensive theory of functional calculus. We develop that theory more fully in this section in order to define parameterizations of $\tens{A}$ with arbitrarily low complexity.

Consider again an arbitrary $\shift\in\LB{\hilbert}$ Laurent with a Jordan spectral representation given by Eq. \eqref{eq:jordan}. The spectrum of $\shift$, denoted $\spectrum{\shift}$, is the union of the eigenvalues of $\hat{\shift}(\omega)$ for $\omega\in\T$,
\begin{equation}\label{eq:spectrum}
\spectrum{\shift} = \cup_{\omega\in\T} \set{\lambda_k(\omega)}_{k=0}^m.
\end{equation}
Let $U\subset\C$ be an open set such that $\spectrum{\shift}\subset U$ and $\phi:U\to\C$ be a holomorphic function. Then, we define
\begin{equation}\label{eq:holomorphic_func_calc}
\phi(\shift) \defined \frac{1}{2\pi\im} \oint_{\Gamma} \phi(z) \left(z\identity - \shift\right)^{-1} dz
\end{equation}
where $\Gamma\subset U$ is a closed curve that encloses $\spectrum{\shift}$ \cite{dunford1966linear}. Let $\tens{A}=\phi(\shift)$. Then, by combining Eqs. \eqref{eq:A_diagonalized_by_F}, \eqref{eq:jordan}, and \eqref{eq:holomorphic_func_calc}, $\tens{A}$ has the following action on $\vect{x}\in\hilbert$:
\begin{equation}\label{eq:action_func_calc}
\left( \tens{A} \vect{x} \right)[t] = \sum_{k=0}^m \left( \iF \left[ \left(\phi\circ\lambda_k\right)(\omega) \proj_k(\omega) + \left(\phi'\circ\lambda_k\right)(\omega) \nil_k(\omega) \right] \cdot \hat{\vect{x}}(\omega) \right)[t]
\end{equation}
where $\circ$ is the operation of composition. Consequently, $\tens{A}\in\LB{\hilbert}$ satisfies Eq. \eqref{eq:covariance}. Moreover, this approach offers a controlled learning complexity. There is $\Oh{1}$ parameter, $\phi:U\to\C$, a holomorphic function on $U\subset\C$. 

As in Defferrard \etal \cite{defferrard_convolutional_2016}, $\phi$ could be a polynomial of degree $0\le p<n$, a parameterization of $\tens{A}$ with $p$ parameters, but it could also be any other holomorphic function on $U$ with arbitrarily few parameters. The class of holomorphic functions on $U\subset\C$ includes the polynomials on $\C$. It is not even necessary that $U\subset\C$ be a simply connected open set. It can be the finite union of disjoint open sets $U=\cup_{j=0}^m U_j$, and $\phi$ need only be holomorphic on the restriction to each $U_j$ with $\Gamma=\cup_{j=0}^m \Gamma_j$ and $\Gamma_j\subset U_j$. This means that we can define holomorphic functions $\phi:\C\to\C$ such that $\phi(z;\alpha,\beta)=\earg{\alpha z + \beta}$ for $z\in U_0$, $\phi(z;\gamma)=+\sqrt{z-\gamma}$ for $z\in U_1$ (assuming $U_1\cap\{0\}=\emptyset$), and $\phi(z)=0$ for $z\in(U_0\cup U_1)^c$ for a total of three learnable parameters $\alpha,\beta,\gamma\in\C$. The parameterization of $\tens{A}$ using functional calculus can be chosen to be of arbitrarily low complexity. 

\section{Example}\label{sec:example}

In this section, we compare the proposed approach to the factor graph model for time-varying graph signals proposed in Sandryhaila and Moura \cite{sandryhaila_big_2014}, an approach primarily motivated by efficient numerical implementation. We choose an example group generator for which we can illustrate analytically the difference in approach. We highlight two advantages, the richness of the graphical model and the separation of the spectrum. 

Consider $\shift\in\LB{\hilbert}$ with kernel function $\kernel$ given by
\begin{equation}\label{eq:s}
\begin{matrix}
\kernel_0 &= \begin{bmatrix}
0 & -1 \\
-1 & 0
\end{bmatrix}
\quad
\kernel_1 &= \begin{bmatrix}
\frac{2}{5} & 0 \\
0 & \frac{2}{5}
\end{bmatrix} 
\\
\kernel_2 &= \begin{bmatrix}
0 & 0 \\
\frac{4}{5} & 0
\end{bmatrix}
\quad
\kernel_3 &= \begin{bmatrix}
0 & \frac{3}{5} \\
0 & 0
\end{bmatrix}
\end{matrix}
\end{equation}
and $\kernel_t=\tens{0}$ otherwise. By Eq. \eqref{eq:A_spectral}, this yields a frequency representation,
\begin{equation}\label{eq:s_spectral}
\hat{\shift}(\omega) = \begin{bmatrix}
\frac{2}{5}\earg{2\pi\im\omega} & -1 + \frac{3}{5}\earg{6\pi\im\omega} \\
-1 + \frac{4}{5}\earg{4\pi\im\omega} & \frac{2}{5}\earg{2\pi\im\omega}
\end{bmatrix},
\end{equation}
a holomorphic matrix-valued function for $\omega\in\T$. $\hat{\shift}(\omega)$ has eigenvalues
\begin{equation}\label{eq:s_eigs}
\lambda_\pm(\omega) = \frac{2}{5}\earg{2\pi\im\omega} \pm \sqrt{\left(1-\frac{3}{5}\earg{6\pi\im\omega}\right)\left(1-\frac{4}{5}\earg{4\pi\im\omega}\right)},
\end{equation}
%
projections
\begin{equation}\label{eq:s_proj}
\proj_\pm(\omega) = \frac{1}{2} \begin{bmatrix}
1 & \pm \sqrt{\frac{5-4\earg{4\pi\im\omega}}{5-3\earg{6\pi\im\omega}}}\\
\pm \sqrt{\frac{5-3\earg{6\pi\im\omega}}{5-4\earg{4\pi\im\omega}}} & 1
\end{bmatrix},
\end{equation}
and nilpotents $\nil_\pm(\omega)=\tens{0}$.

Alternatively, we could follow the approach proposed in Sandryhaila and Moura \cite{sandryhaila_big_2014}, in which time-varying graph signals are modeled with the Cartesian graph product (Eq. (25) of \cite{sandryhaila_big_2014}) of the cyclic shift (Eq. (3) of \cite{sandryhaila_big_2014}) and the weighted adjacency matrix on $\graph$ for which we will use $\tens{W}=\sum_{t\in\Z}\kernel_t$. Then, we define an associated group generator
\begin{equation}\label{eq:s0}
\shift_0 = \begin{bmatrix}
\\
\ddots & \\
& 1 & \\
& & \ddots & 
\end{bmatrix} 
\otimes \identity + \identity \otimes \left( \sum_{t\in\Z}\kernel_t \right),
\end{equation}
which by Eq. \eqref{eq:A_spectral}, yields a frequency representation,
\begin{equation}\label{eq:s0_spectral}
\hat{\shift}_0(\omega) = \begin{bmatrix}
\earg{2\pi\im\omega} + \frac{2}{5} & -\frac{2}{5} \\
-\frac{1}{5} & \earg{2\pi\im\omega} + \frac{2}{5}
\end{bmatrix}.
\end{equation}
$\hat{\shift}_0(\omega)$ has eigenvalues
\begin{equation}\label{eq:s0_eigs}
\left(\lambda_0\right)_\pm(\omega) = \earg{2\pi\im\omega} + \frac{2\pm\sqrt{2}}{5},
\end{equation}
%
projections
\begin{equation}\label{eq:s0_proj}
\left(\proj_0\right)_\pm(\omega) = \left(\proj_0\right)_\pm = \frac{1}{2} \begin{bmatrix}
1 & \mp \frac{2}{\sqrt{2}} \\
\mp \frac{1}{\sqrt{2}} & 1
\end{bmatrix},
\end{equation}
and nilpotents $\nil_\pm(\omega)=\tens{0}$. 

\begin{figure}[!ht]
\begin{minipage}[b]{.48\linewidth}
 \centering
 \centerline{\includegraphics[width=\linewidth]{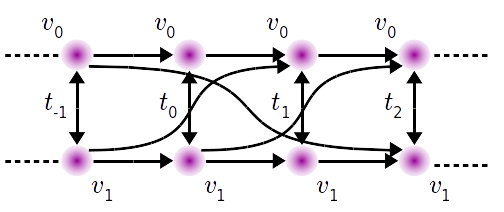}}
 \centerline{(a) Edges of $\shift$}
\end{minipage}
\hfill
\begin{minipage}[b]{0.48\linewidth}
 \centering
 \centerline{\includegraphics[width=\linewidth]{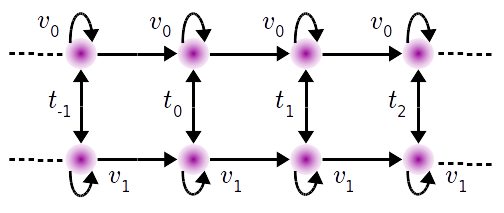}}
 \centerline{(b) Edges of $\shift_0$}
\end{minipage}
\caption{A visual depiction of the weighted edges of $\shift$ and $\shift_0$. $\shift$ offers a richer model of the temporal interaction between nodes. Edges connect nodes across zero, one, two, and three time steps, whereas $\shift_0$ has only temporal interaction across one time step.}
\label{fig:graphs}
\end{figure}

Note the alternative graphical models in Fig. \ref{fig:graphs}. The greater flexibility of $\shift$ can facilitate modeling of more complex systems since the edges of $\shift$ offer more pathways by which nodes can interact to influence the system behavior. Importantly, for even this simple example, $\shift$ does in fact yield more complex behaviors as is shown in the following.

\begin{figure}[!ht]
\begin{minipage}[b]{.48\linewidth}
 \centering
 \centerline{\includegraphics[width=\linewidth]{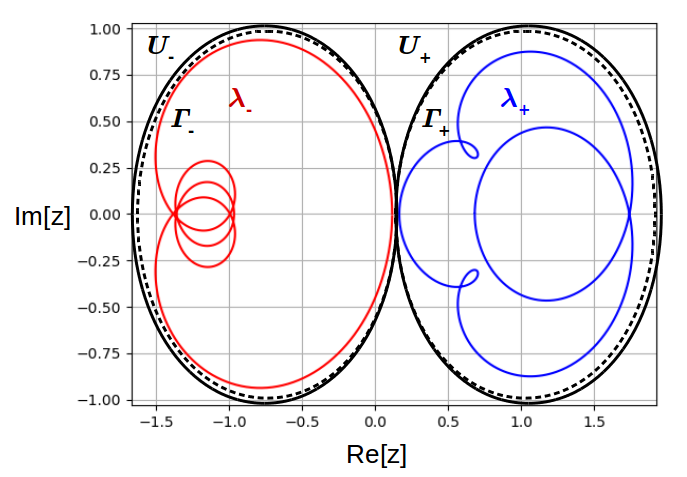}}
 \centerline{(a) Spectrum of $\shift$}
\end{minipage}
\hfill
\begin{minipage}[b]{0.48\linewidth}
 \centering
 \centerline{\includegraphics[width=\linewidth]{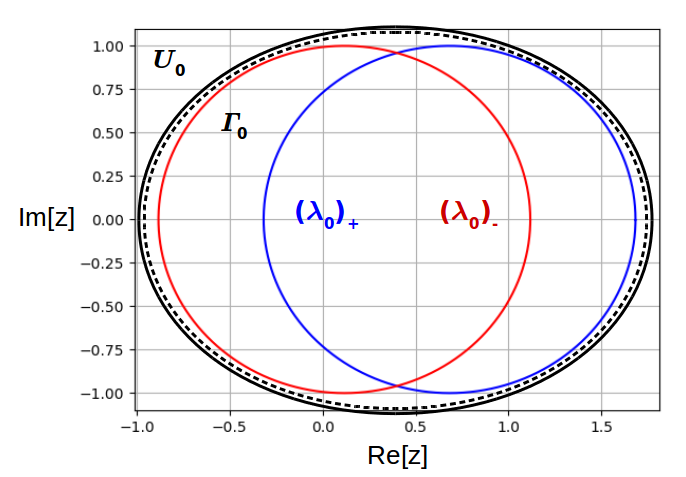}}
 \centerline{(b) Spectrum of $\shift_0$}
\end{minipage}
\caption{The spectrum of $\shift$ and $\shift_0$. (a) $\cup_{\omega\in\T}\lambda_+(\omega)$ (blue) and $\cup_{\omega\in\T}\lambda_-(\omega)$ (red) compose $\spectrum{\shift}$. These are holomorphic functions of $\omega\in\T$ for which we can define an open set $U=U_+\cup U_-$ such that $\cup_{\omega\in\T}\lambda_+(\omega)\subset U_+$ and $\cup_{\omega\in\T}\lambda_-(\omega)\subset U_-$ and $U_+\cap U_-=\emptyset$. (b) $\cup_{\omega\in\T}(\lambda_0)_+(\omega)$ (blue) and $\cup_{\omega\in\T}(\lambda_0)_-(\omega)$ (red) compose $\spectrum{\shift_0}$, also holomorphic functions of $\omega\in\T$. However, the spectrum is not separable, and we must define an open set $U_0\subset\C$ such that $\spectrum{\shift_0}\subset U_0$.}
\label{fig:eigenvalues}
\end{figure}

Let $\phi:\C\to\C$ be a circular complex Gaussian,
\begin{equation}\label{eq:complex_gaussian}
\phi(z; \mu, \sigma, U) = \begin{cases} \frac{1}{2\pi\sigma} \exparg{ -\frac{1}{\sigma^2} \abs{z-\mu}^2 } & z\in U \\ 0 & o.w. \end{cases}
\end{equation}
for $U\subset\C$, $\mu\in \C$ and $\sigma\in\R$ \cite{picinbono_second-order_1996}. Define $U=U_+\cup U_-$ as in Fig. \ref{fig:eigenvalues}. Then, $\tens{A}=\phi(\shift; \mu, \sigma, U_+)\in\LB{\hilbert}$ satisfies Eq. \eqref{eq:covariance}. Now, consider the action of $\tens{A}$ in the limit as $\sigma\to0$ and $\mu\subset\cup_{\omega\in\T} \lambda_+(\omega)$. Then, using Eq. \eqref{eq:action_func_calc}, the following limit holds in the distribution sense:
\begin{equation}\label{eq:s_ideal_bandpass}
\lim_{\sigma\to0} \left( \tens{A} \vect{x} \right)[t] = \earg{-2\pi\im\mu t} \proj_+(\mu) \hat{\vect{x}}(\mu).
\end{equation}
Here, we can understand $\tens{A}$ as implementing an ideal bandpass in time and space.  The projections,\\ $\cup_{\omega\in\T}\set{\proj_+(\omega),\proj_-(\omega)}$, of Eq. \eqref{eq:s_proj} define the invariant subspaces of $\shift$. For each element $\lambda_0\in\spectrum{\shift}$, there is an associated subspace in $\hilbert$. Eq. \eqref{eq:s_ideal_bandpass} shows that these modes defined by the range of $\proj_\pm(\mu)$ for $\mu\in\T$ manifest significantly different behaviors for different frequencies. 

Now consider $\shift_0$. Define $U_0$ as in Fig. \ref{fig:eigenvalues} and let $\tens{A}_0=\phi(\shift_0;\mu,\sigma,U_0)$ for $\mu\in\cup_{\omega\in\T}(\lambda_0)_+(\omega)$. $\tens{A}_0$ satisfies Eq. \eqref{eq:covariance} for $\shift_0$. We can consider an ideal bandpass similar to Eq. \eqref{eq:s_ideal_bandpass}, and again, it would hold in the distribution sense,
\begin{equation}\label{eq:s0_ideal_bandpass}
\lim_{\sigma\to0} \left(\tens{A}\vect{x}\right)[t] = \earg{-2\pi\im\mu t} (\proj_0)_+ \hat{\vect{x}}(\mu).
\end{equation}

However, notice that in Eqs. \eqref{eq:s0_proj} and \eqref{eq:s0_ideal_bandpass}, the projections of $\shift_0$, $\set{(\proj_0)_+,(\proj_0)_-}$, are not functions of $\omega$. The frequency-dependent modes of the network are somehow lost in the factor graph model. Regardless of frequency, the behavior at the nodes will be the same. This has important implications for learning informative representations of complex systems which display \eg cross-frequency coupling. \textit{A priori}, we may not know the important behaviors or interactions of the network for tasks such as discrimination, regression, or compression. Having a powerful and flexible model which can learn the relevant representations is then exceedingly important.

Compare now the spectra of $\shift$ and $\shift_0$ in Fig. \ref{fig:eigenvalues}. That the spectra are well separated for $\shift$ is important for the application of the functional calculus. For $\shift$, We can define a different holomorphic function restricted to each separable compact set of $\spectrum{\shift}$ as described in Sec. \ref{sec:covariant_func_calc}. For instance, we can define $\tens{A}=\phi(\shift;\mu_+,\sigma,U_+)+\phi(\shift;\mu_-,\sigma,U_-)$ to learn simultaneous projections associated with $U_+$ and $U_-$ to find rich inter-relationships between modes of the stochastic process on the graph with two learnable parameters, $\set{\mu_+,\mu_-}$. For $\shift_0$, due to the lack of separation of the spectra, any holomorphic function must be applied uniformly on the entire spectrum.

\section{Conclusion}\label{sec:conclusion}

We have proposed a theoretical framework for learning robust representations of stochastic processes on directed graphs with arbitrarily low complexity. We applied that theory to an example problem that illustrates the advantages of the proposed approach over factor graph models such as those proposed in Sandryhaila and Moura \cite{sandryhaila_big_2014}. Specifically, the proposed framework yields greater model expressiveness. Importantly, the framework can be implmented with $\Oh{1}$ learning complexity. Future work will incorporate the proposed theory into a graph convolutional network and demonstrate its advantage on real-world applications such as social network analysis or brain imaging.

\bibliographystyle{plain}
\bibliography{references}

\end{document}